\documentclass[aps,floatfix,nofootinbib,preprintnumbers,amsmath, amssymb,showkeys,preprint]{revtex4-1}

\usepackage{epsf,epsfig,subfigure,graphicx,amsmath,amssymb}
\usepackage{color}

\def\lsim{\mathrel{\rlap{\lower4pt\hbox{\hskip1pt$\sim$}}
    \raise1pt\hbox{$<$}}}         
\def\gsim{\mathrel{\rlap{\lower4pt\hbox{\hskip1pt$\sim$}}
    \raise1pt\hbox{$>$}}}         

\begin{document}

\title{\bf Galactic center GeV gamma-ray excess\\
 from dark matter with gauged lepton numbers }

\author{
Jongkuk Kim$^{(1)}$\footnote{jongkukkim@skku.edu},
Jong-Chul Park$^{(1, 2)}$\footnote{jcpark@cnu.ac.kr}, 
and Seong Chan Park$^{(3, 4)}$\footnote{sc.park@yonsei.ac.kr}
}

\affiliation{
$^{(1)}$ Department of Physics, Sungkyunkwan University, Suwon 440-746, Korea\\
$^{(2)}$ Department of Physics, Chungnam National University, Daejeon 305-764, Korea\\
$^{(3)}$Department of Physics \& IPAP, Yonsei University, Seoul 120-749, Korea\\
$^{(4)}$ Korea Institute for Advanced Study, Seoul 130-722, Korea
}

\vspace{1.0cm}
\begin{abstract}
The recently observed excess in gamma-ray signal near the Galactic center suggests that dark matter particles may annihilate into charged fermions that produce gamma-ray to be observed.
In this paper, we consider a leptonic dark matter, which annihilates into the standard model leptons,  $\mu^+ \mu^-$ and $\tau^+ \tau^-$, by the interaction of the gauged lepton number ${\rm U(1)}_{L_\mu-L_\tau}$ and  fits the observed excess.
Interestingly, the necessary annihilation cross section for the observed gamma-ray flux provides a good fit to  the value for the relic abundance of dark matter.
We identify the preferred parameter space of the model after taking the existing experimental constraints from the precision measurements including the muon $(g-2)$, tau decay, neutrino trident production, dark matter direct detection, LHC, and LEP experiments.
\end{abstract}


\maketitle

\section{Introduction}

The dark matter (DM) problem is one of the pressing issues in particle physics and cosmology.
While the existence of DM has been firmly established through various observations of its gravitational effects on multiple scales, its microscopic nature still remains unknown \cite{Jungman:1995df}.
This situation stimulates a variety of DM searches including the direct detection of dark matter scattering off detector materials, the detection of indirect signals from the dark matter annihilation or decay, and the collider searches of missing energy signatures due to the produced dark matter particles.
Of particular, we notice that the new cosmic-ray detection experiments, such as PAMELA~\cite{Adriani:2008zr}, AMS-02~\cite{Aguilar:2013qda}, and Fermi-LAT~\cite{FermiLAT:2011ab}, based on satellites reach unprecedented sensitivity to the cosmic-ray signals, which leads better chance to get the indirect information of dark matter properties.

An intriguing observation was made using the public data of the Fermi Large Area Telescope (Fermi-LAT) by
Hooper {\it et. al.} and also other independent groups~\cite{Hooper1, Hooper2, Hooper3, Abazajian:2012pn, Hooper4, Gordon:2013vta, Huang:2013pda, Abazajian:2014fta, HooperNew, Silk, Calore:2014xka, Calore:2014nla}: a gamma-ray excess at $E_\gamma \approx \mathcal{O}({\rm GeV})$ coming from the Galactic center (GC) is found.
In the analyses, it is claimed that the gamma-ray excess spectrum is in good agreement with the emission expected from DM annihilation into standard model (SM) charged particles.\footnote{In Ref.~\cite{Kim:2015usa}, the authors proposed a new mechanism naturally inducing a continuum bump signature in cosmic $\gamma$-ray measurements even with a particle directly decay into two photons, introducing Energy Peak idea together with the postulate of a generic dark sector~\cite{Belanger:2011ww}.}
The GeV excess is well fitted by a DM particle with a mass of $m_{\rm DM } \approx 30-40$ GeV annihilating to $b\overline{b}$ with an annihilation cross section of $\langle\sigma v\rangle \approx 2 \times 10^{-26} {\rm cm}^3/{\rm s}$~\cite{HooperNew, Calore:2014nla}.\footnote{We note that recent observation of AMS-02~\cite{AMS2015} has started to exclude the $\chi\chi \to b\bar{b}$ dominant DM explanation of relic abundance~\cite{Giesen:2015ufa}.}
Silk {\it et. al.} pointed out that contributions of the diffuse photon emissions from primary and secondary electrons produced in DM annihilation processes are significant, especially for leptonic final states ($\ell \bar \ell$)~\cite{Silk}.
It is also noticed that with the inverse Compton scattering (ICS) and bremsstrahlung contributions from electrons, annihilations of DM particles with a mass of $m_{\rm DM } \approx 10$ GeV into $\ell \bar \ell$ provide a good fit with an annihilation cross section of $\langle\sigma v\rangle \approx (1-2) \times 10^{-26} {\rm cm}^3/{\rm s}$~\cite{Silk}.
The $b\bar{b}$ final state may be understood by Higgs portal type DM models and studied by several authors \cite{Okada:2013bna, Alves:2014yha, Ipek:2014gua, Basak:2014sza} but a model for the leptonic explanation based on leptophilic DM is relatively less studied for the GeV excess. Here we explore a leptophilic model with the DM mass $m_{\rm DM } \approx 10$ GeV.

In the heavier mass domain, $M_{\rm DM}\gsim 100$ GeV, leptophilic DM models have attracted a lot of attention (see e.g.~\cite{Chun:2008by}.) due to recent observation of excessive cosmic-ray positron fraction by the PAMELA, Fermi-LAT, and AMS-02 experiments, but lack of excess in the anti-proton fraction~\cite{Adriani:2010rc}.
In building the leptophilic DM model, it is attractive to gauge the differences in lepton numbers: U(1)$_{L_e-L_\mu}$, U(1)$_{L_\mu-L_\tau}$, and U(1)$_{L_\tau-L_e}$.  These symmetries are anomaly free without extending the SM particle contents~\cite{He:1990pn, He:1991qd, Foot:1990mn}.\footnote{The other anomaly free choice is U(1)$_{B-L}$, but it does not provide lepton specific interactions.}
Leptophilic DM models with a U(1)$_{L_i-L_j}$ gauge symmetry for the positron excess were studied in Refs.~\cite{Baek:2008nz, Bi:2009uj, Das:2013jca}.
In our analysis, we take U(1)$_{L_\mu-L_\tau}$ symmetry  for the GeV gamma-ray excess since models with U(1)$_{L_e-L_\mu}$, and U(1)$_{L_\tau-L_e}$ are stringently limited by existing cosmic-ray positron measurements in low energy~\cite{Kong:2014haa}.

It should be also noticed that astrophysical uncertainty in gamma-rays from around the GC including modeling of background emission in the inner galaxy is still big.
Moreover, millisecond pulsars~\cite{Hooper2, Hooper3, Abazajian:2012pn, Gordon:2013vta, Abazajian:2014fta, Abazajian:2010zy} and pions from the collision of cosmic-rays with gas~\cite{Hooper2, Hooper3, Abazajian:2012pn, Gordon:2013vta} can contribute to the GeV scale gamma-ray and have been proposed as alternative explanations of the GeV gamma-ray excess even though the spectral shape from millisecond pulsars looks too soft at the sub-GeV energy range to account for the observed GeV excess spectrum~\cite{Hooper:2013nhl}.
Also the morphological feature of the observed excess is extended to more than $\sim 10^\circ$ from the GC beyond the boundary of the central stellar cluster which could include numerous millisecond pulsars~\cite{HooperNew}, and observed distributions of gas seem to provide a poor fit to the spatial distribution of the signal~\cite{Linden:2012iv, Macias:2013vya, HooperNew}.

The contents of the paper is as follows.
In Section II, we explain the leptophilic DM model in detail and present dominant annihilation channels.
The model parameter space for the observed DM thermal relic abundance and the GeV gamma-ray excess is clarified.
In Section III, we discuss the existing experimental constrains on the same parameter space, then conclude in Section IV.

\newpage
\section{The model, relic abundance, and Fermi-LAT GeV excess}

\subsection{The model}

We extend the SM:
\begin{itemize}
\item by extending the gauge symmetry with U(1)$_{L_\mu-L_\tau}$,
\item by introducing a new Dirac fermion $\psi$, which is identified as dark matter.
\end{itemize}
The charge assignment for the SM fermions and the new fermion regarding the $L_\mu-L_\tau$ symmetry is given in Table~\ref{charges}.
The muon-leptons and anti-tau leptons are $(+1)$, tau-leptons and anti-muon leptons are $(-1)$ and the new fermion has a charge $Q_\psi'$. We take a universal gauge coupling constant $g'$ for $Z'$ interactions.

\vspace{0.5 cm}
\begin{table}[h]
\begin{center}
\begin{tabular}{|c||c|c|c|c|}
\hline
~particle~ & $\psi$ & ~$L_\mu=(\nu_{\mu L},\mu_L),~\mu_R,~\nu_{\mu R}$~ & ~$L_3=(\nu_{\tau L},\tau_L),~\tau_R,~\nu_{\tau R}$~ & ~others~\\
\hline
charge & ~$Q'_\psi$~ & $+1$ & $-1$  & 0\\
\hline
\end{tabular}
\end{center}
\caption{Charges under the $L_\mu-L_\tau$ gauge symmetry.}
\label{charges}
\end{table}

For the (spontaneously broken) extended gauge symmetry, we associate a new vector boson $Z'$ with an undetermined mass $m_{Z'}$. The model Lagrangian is written as follows:
\begin{eqnarray} \label{Lagrangian}
{\cal L}~ \supset~ {\cal L}_{SM} &-&
{1\over4} Z'_{\alpha\beta} Z'^{\alpha\beta} + {1\over2} m_{Z'}^2 Z'_\alpha Z'^\alpha
+i\overline{\psi}\gamma_\alpha\partial^\alpha\psi - m_\psi \overline{\psi}\psi \nonumber\\
&+& g' Q'_\psi Z'_\alpha \overline{\psi}\gamma^\alpha\psi + g' Z'_\alpha \sum_{f=\mu, \tau, \nu_\mu, \nu_\tau} Q'_f \overline{f}\gamma^\alpha f\,,
\end{eqnarray}
where $Q'_{\psi, f}$ are U(1)$_{L_\mu-L_\tau}$ charges of the DM and a SM fermion $f$, respectively given in  Table~\ref{charges}.
In our study, the DM mass $m_\psi$ is taken to be 10 GeV to fit the GeV excess as suggested in Ref.~\cite{Silk} (see also Ref.~\cite{Kyae:2013qna}).

\begin{figure}[h]
\begin{center}
\includegraphics[width=0.45\linewidth]{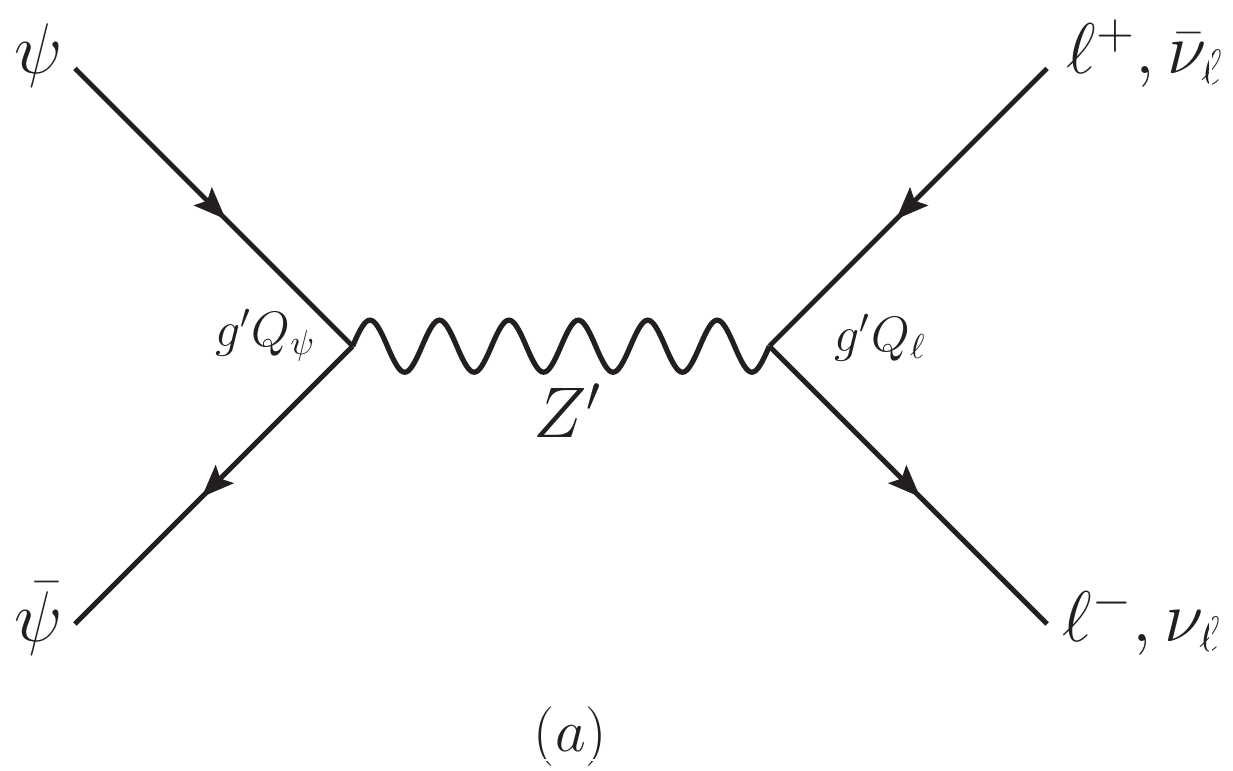}
\hspace*{0.5cm}
\includegraphics[width=0.33\linewidth]{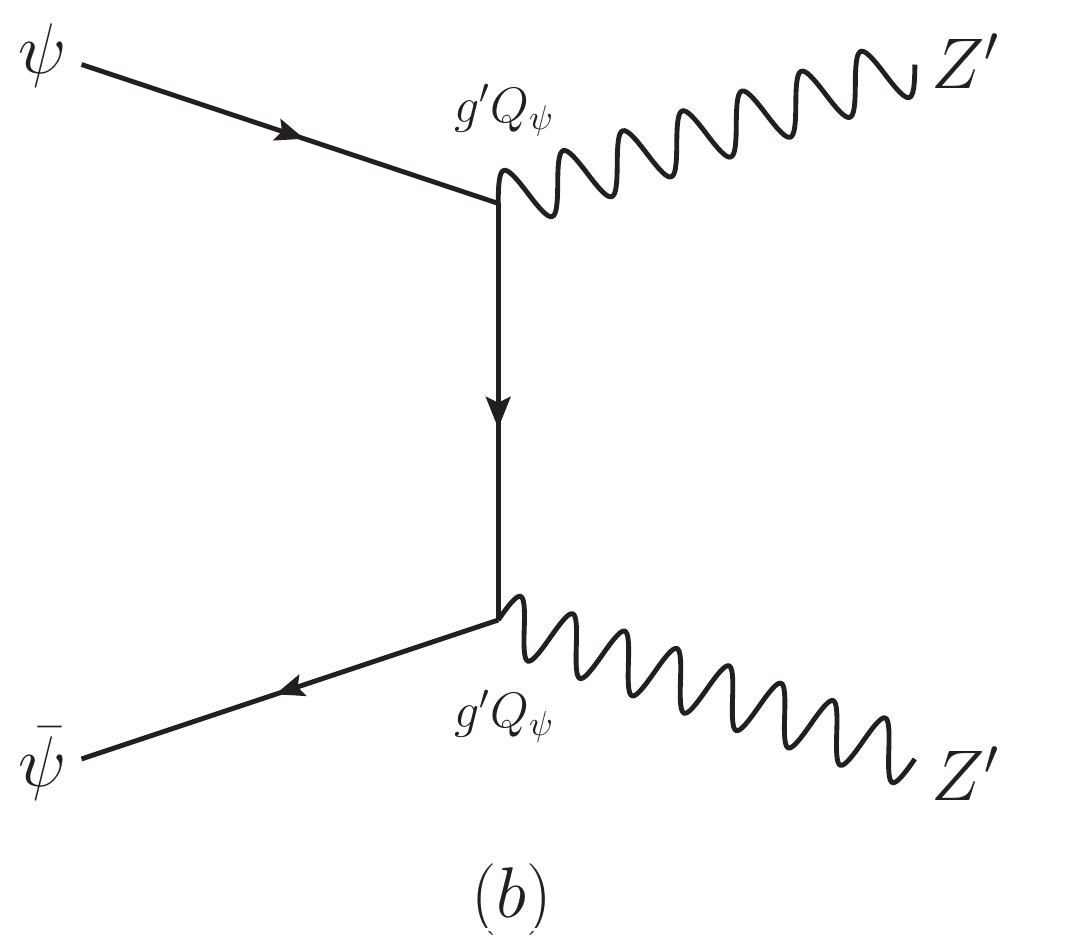}
\end{center}
\vspace*{-0.7cm}
\caption{Dominant DM annihilation channels: (a) $s$-channel annihilations into leptons ($\ell=\mu$, $\tau$) through a $Z'$ boson exchange and (b) $t$-channel annihilation into a pair of $Z'$ bosons.
}
\label{Fig-Annihilation}
\end{figure}

The $\psi$ particle is neutral under the SM gauge interactions but its presence is seen by $L_\mu-L_\tau$ interactions.
The gauge interaction allows an early time thermal equilibrium with the SM particles and the standard freeze-out took place at $T\sim m_\psi/20$  through the  the dominant annihilation channels:
\begin{eqnarray}
&& \psi\overline{\psi} ~(\to Z'^{(*)}) \to  \ell^+\ell^-,~\nu_\ell\overline{\nu}_\ell\\
&& \psi\overline{\psi} \to Z' Z'\,,
\end{eqnarray}
where $\ell=\mu,~\tau$.
The corresponding Feynman diagrams are depicted in Fig.~\ref{Fig-Annihilation}.
The DM annihilation into a $Z'$ pair is kinematically allowed only when $m_\psi > m_{Z'}$.

%
%

The leading order DM annihilation cross sections are given by
\begin{eqnarray}
\langle\sigma v \rangle_{\psi\overline{\psi}\to\ell\overline{\ell}}
& \simeq &
\frac{g^{\prime 4} Q^{\prime 2}_\psi}{2\pi}
\frac{m^2_\ell +2m^2_\psi}{\left( m^2_{Z^\prime}-4m^2_\psi \right)^2+ m^2_{Z^\prime}\Gamma^2_{Z^\prime}}\,
\sqrt{1-\frac{m^2_\ell}{m^2_\psi}} +\mathcal{O}(v^2)\,, \label{Annihilation_ll} \\
\langle\sigma v \rangle_{\psi\overline{\psi}\to Z^\prime Z^\prime}
& \simeq &
\frac{g^{\prime 4} Q^{\prime 2}_\psi}{4\pi}
\frac{m^2_\psi -m^2_{Z^\prime}}{\left( m^2_{Z^\prime} -4m^2_\psi \right)^2}\, \sqrt{1-\frac{m^2_{Z^\prime}}{m^2_\psi}} +\mathcal{O}(v^2)\,, \label{Annihilation_ZZ}
\end{eqnarray}
where $\ell=\mu,~\tau,~\nu_\mu$, and $\nu_\tau$.
The decay width of the $Z'$ boson is given by
\begin{eqnarray}
\Gamma_{Z'} \simeq \sum_{\ell=\mu, \tau, \nu_\mu, \nu_\tau} &&
\frac{g'^2}{12 \pi m_{Z'}}\left( m^2_{Z'} +2m^2_\ell \right) \sqrt{1-\frac{4m^2_\ell}{m^2_{Z'}}} ~\theta\left(m_{Z'} - 2m_\ell \right) \nonumber\\
 +&& \frac{g^{'2}Q'^{2}_\psi}{12 \pi m_{Z'}}\left( m^2_{Z'} +2m^2_\psi \right) \sqrt{1-\frac{4m^2_\psi}{m^2_{Z'}}} ~\theta\left(m_{Z'} -2m_\psi\right)\,,
\end{eqnarray}
where $\theta$ is the unit step function.

\subsection{Relic abundance}

%
\begin{figure}[h]
	\begin{center}
		\includegraphics[width=0.70\linewidth]{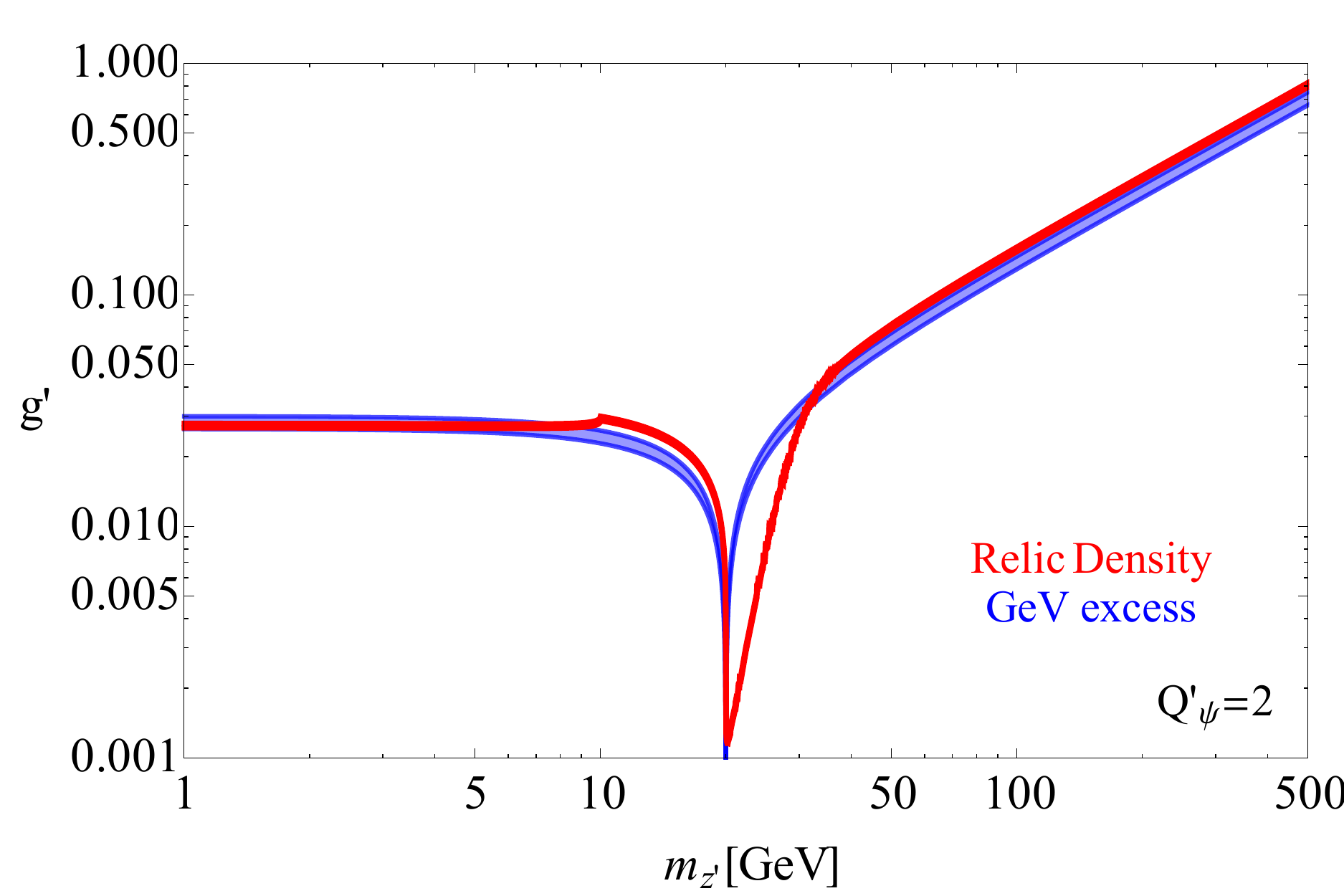}
	\end{center}
	\vspace*{-0.7cm}
	\caption{Preferred parameter regions in the $m_{Z'}-g'$ plane for $Q'_\psi=2$.
In the narrow red band, the relic density of DM $\psi$ is in the range of $0.11 < \Omega_{\rm DM}h^2 < 0.13$.
In the blue band, the annihilation cross section into $\mu^+\mu^-$ and $\tau^+\tau^-$ satisfies $\langle\sigma v \rangle_{\psi\overline{\psi}\to \mu^+\mu^-, \tau^+\tau^-} \approx
(0.95-1.49)
\times 10^{-26} {\rm cm}^3/{\rm s}$, which is required to fit the GeV gamma-ray excess.
	}
	\label{Fig-Relic}
\end{figure}
%

Taking the DM relic density $0.11 < \Omega_{\rm DM}h^2 < 0.13$~\cite{Planck2015}, we found the preferred parameter space for $\psi$ dark matter in $m_{Z'}-g'$ plane for $Q'_\psi=2$ in Figure~\ref{Fig-Relic}.
The plots for other values of $Q_\psi'$ are also given later.
The ballpark range is $1 <m_{Z'} [{\rm GeV}] < 500$ and $0.001 < g'<1.0$ as a reasonable choice within the perturbative regime.
Notably, the dip structure appears around $m_{Z'}\simeq 2m_\psi = 20$ GeV due to the resonance in the $s$-channel annihilation into leptons mediated by the $Z'$ gauge boson.
In calculating the thermal average of DM annihilation cross section for the relic abundance, we take the  the non-negligible effect of DM kinetic energy near the resonance pole, $m_{Z'}\simeq 2m_\psi = 20$ GeV as explained in Ref.~\cite{Griest:1990kh}.
The result is shown in Fig.~\ref{Fig-Relic}.

\subsection{Fermi-LAT GeV excess}

%
\begin{figure}[h]
	\begin{center}
		\includegraphics[width=0.70\linewidth]{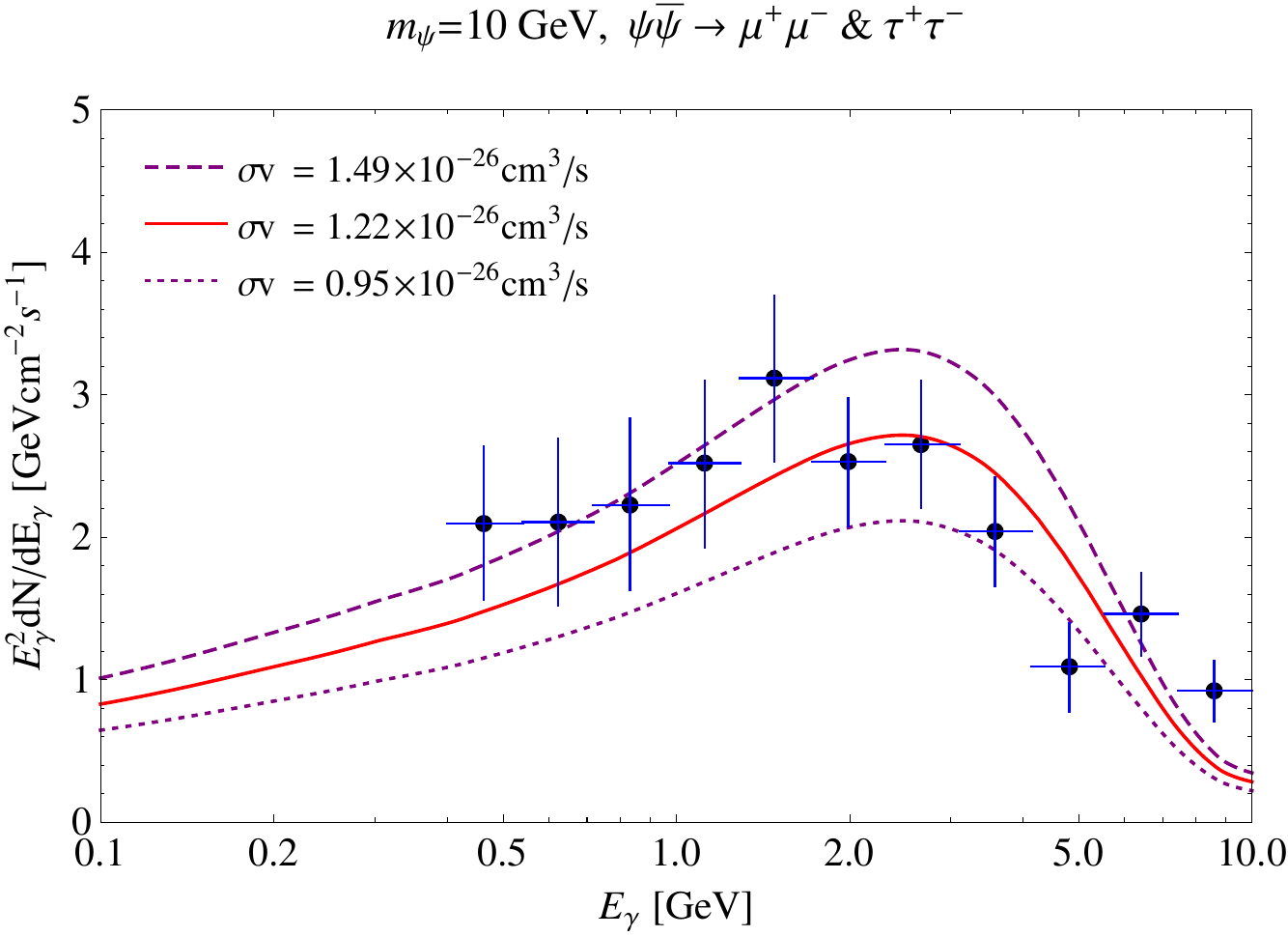}
	\end{center}
	\vspace*{-0.7cm}
	\caption{
Fits to the GC GeV $\gamma$-ray excess for 10 GeV DM annihilating into $\mu^+\mu^-$ and $\tau^+\tau^-$ with branching ratios of $\mu^+\mu^-:\tau^+\tau^-=1:1$.
The best fit is obtained with $\langle\sigma v \rangle \approx 1.22 \times 10^{-26} {\rm cm}^3/{\rm s}$, which is plotted as a red line.
Upper and lower fits corresponding to a p-value greater than $10^{-3}$ are presented as purple dashed and dotted curves, respectively.
The black points with blue error bars are the data points extracted from Ref.~\cite{Silk}.
}
	\label{Fig-FermiFitting}
\end{figure}
%

We conduct the fit of our model, $\mu^+\mu^-:\tau^+\tau^-=1:1$, to the observed spectrum of the GC GeV $\gamma$-ray excess.
Our best fit is obtained for $\langle\sigma v \rangle_{\psi\overline{\psi}\to \mu^+\mu^-, \tau^+\tau^-} \approx 1.22 \times 10^{-26} {\rm cm}^3/{\rm s}$ with $\chi^2=19.22$.  Our result can be compared with the results in Ref.~\cite{Silk}, where  the best fits are $\langle\sigma v \rangle \approx 0.86 \times 10^{-26} {\rm cm}^3/{\rm s}$ with $\chi^2=10.21$ for the democratic leptonic final state and $\langle\sigma v \rangle \approx 1.42 \times 10^{-26} {\rm cm}^3/{\rm s}$ with $\chi^2=14.22$ for branching ratios of $\mu^+\mu^-:\tau^+\tau^-=2:1$.
Accepting the $\chi^2 < 29.6$, which provides a p-value larger than $10^{-3}$ for 10 degrees of freedom (i.e. 11 data points and one fitting parameter $\langle\sigma v \rangle$),
we found the preferred interval $\langle\sigma v \rangle = (0.95-1.49) \times 10^{-26} {\rm cm}^3/{\rm s}$.
In Fig.~\ref{Fig-FermiFitting}, we plot our best fit as a red line with an interval  corresponding to a p-value of $10^{-3}$.
The data points are presented by black dots and their error bars are represented by blue lines.
As can be seen in Fig.~\ref{Fig-Relic}, the plot for the right relic abundance reproduces successful GeV excess in the GC as was originally observed in~\cite{Hooper1, Hooper2} and also in~\cite{Silk} for leptonic annihilations taking into account the contributions by the ICS and bremsstrahlung with the annihilation cross section of $\langle\sigma v \rangle \approx (0.95-1.49) \times 10^{-26} {\rm cm}^3/{\rm s}$ for the preferred mass range near $10$ GeV.
{
In Fig.~\ref{Fig-Relic}, for the case with $Q'_\psi=2$, the parameter space fitting both the relic abundance and the GC GeV excess lies in three regions $(m_{Z'} [{\rm GeV}], ~g') =(\lesssim 9.6, ~0.027)$, $(19.7-20.3, ~0.006-0.0012)$, and $(30-42, ~0.028-0.056)$.
The parameter space would be slightly changed with different values of $Q'_\psi$: e.g. for $Q'_\psi =0.1$,  $(19.7-20.3, ~0.026-0.0054)$ and $(27-44, ~0.10-0.27)$.
}

\section{Experimental constraints for the preferred parameter space}

We now check whether the preferred parameter space $m_{Z'} \sim \mathcal{O}(10-100)$ GeV and $g'<1$ is still available after taking the relevant experimental constraints from the processes potentially induced by the gauged lepton number interactions:  $(g-2)_\mu$, $\tau$ decay, neutrino trident production, loop-induced DM-nucleon scattering and  leptonically interacting $Z'$ searches at colliders.

\subsection{\bf $(g-2)_\mu$}

\begin{figure}
\begin{center}
\includegraphics[width=0.37\linewidth]{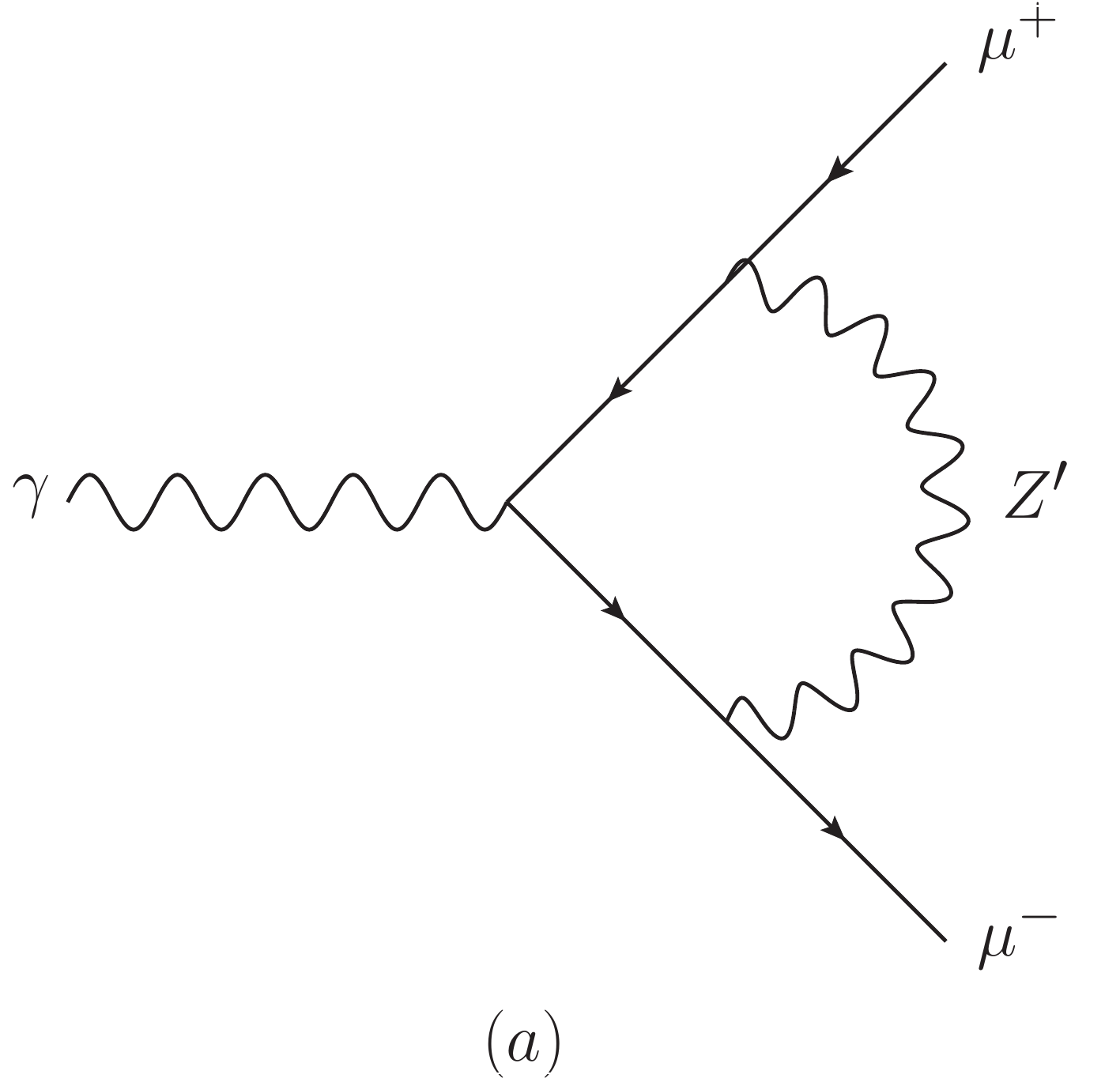}
\hspace*{0.5cm}
\includegraphics[width=0.51\linewidth]{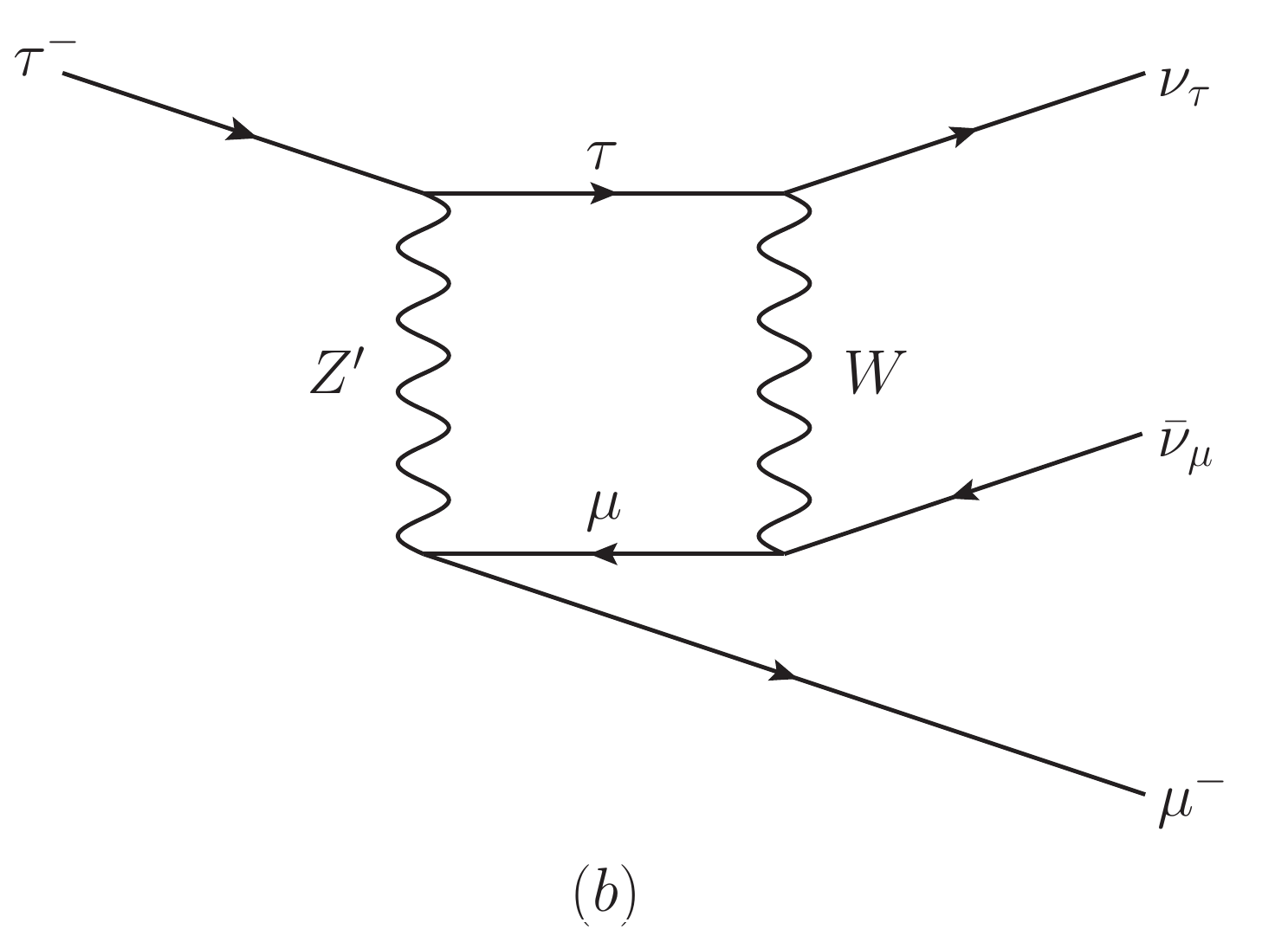}
\end{center}
\vspace*{-0.7cm}
\caption{Feynman diagrams that give a correction to (a) $(g-2)_\mu$ and (b) $\tau \to \mu \nu_\tau \overline{\nu}_\mu$ decay.
}
\label{Fig-g2taudecay}
\end{figure}

The gauged lepton number interaction leads to corrections to the muon anomalous magnetic moment $a_\mu=(g-2)_\mu$ through a Feynman diagram shown in Fig. \ref{Fig-g2taudecay}(a).
The one-loop contribution to $(g-2)_\mu$ is given by \cite{Baek:2001kca, Ma:2001md, Harigaya:2013twa}
\begin{eqnarray}\label{Delta_g-2}
\Delta a_\mu \simeq \frac{g'^2}{12 \pi^2}\, \frac{m_\mu^2}{m_{Z'}^2}\,,
\end{eqnarray}
where we assume that $m_{Z'} \gg m_\mu$, which is indeed valid with $m_{Z'}\sim 10~{\rm GeV}$.
The experimentally measured value and the SM prediction of $(g-2)_\mu$ are respectively given as  \cite{Agashe:2014kda}
\begin{eqnarray}\label{g-2EXP}
a_\mu^{\rm Exp} &=& (11659209.1 \pm 6.3) \times 10^{-10}\,,\\
a_\mu^{\rm SM} &=& (11659180.3 \pm 4.9) \times 10^{-10}\,.
\label{g-2SM}
\end{eqnarray}
Thus, there exists discrepancy between the experimental value and the SM prediction:
\begin{eqnarray}\label{Difference_g-2}
\Delta a_\mu \equiv a_\mu^{\rm Exp} - a_\mu^{\rm SM} = (28.8 \pm 8.0) \times 10^{-10}\,.
\end{eqnarray}
Even though the difference may be a sign of new physics but, more conservatively, we would set an upper bound on the size of the new contribution given in Eq.~(\ref{Delta_g-2}) and find the $2\sigma$ bound line in Fig.~\ref{Fig-result}.

%
%

\subsection{$\tau$ decay}

The gauged lepton number interaction may be seen in the leptonic decay of  tau through the box diagrams such as  Fig.~\ref{Fig-g2taudecay}(b), which could make the branching fraction, ${\rm Br}(\tau \to \mu \nu_\tau \bar{\nu}_\mu)$, larger than what the SM predicts.
It is interesting to notice that the measured value of the tau decay branching fraction to $\mu \nu_\tau \overline{\nu}_\mu$ is indeed slightly larger than what the SM predicted:
the measured values for the branching ratio, ${\rm Br}(\tau \to \mu \nu_\tau \overline{\nu}_\mu)$, and the life time of tau from the PDG~\cite{Agashe:2014kda} are
\begin{eqnarray}\label{taudecayEXP}
{\rm Br}(\tau \to \mu \nu_\tau \overline{\nu}_\mu)|_{\rm PDG} &=& (17.41 \pm 0.04) \%,\\
\tau_\tau|_{\rm PDG} &=& (290.3 \pm 0.5) \times 10^{-15} {\rm s}\,,
\end{eqnarray}
which has  more than $2\sigma$ deviation from  the SM prediction~\cite{Pich:2013lsa, Altmannshofer:2014cfa}:
\begin{eqnarray}\label{Difference_tau}
\frac{{\rm Br}(\tau \to \mu \nu_\tau \overline{\nu}_\mu)}{{\rm Br}(\tau \to \mu \nu_\tau \overline{\nu}_\mu)_{\rm SM}} \simeq 1 + \Delta \quad {\rm with}\quad
\Delta = (7.0 \pm 3.0) \times 10^{-3}\,.
\end{eqnarray}
From the box diagrams with the $Z'$ mediation, the deviation $\Delta$ could be evaluated~\cite{Altmannshofer:2014cfa}:
\begin{eqnarray}\label{Delta_tau}
\Delta = \frac{3 g'^2}{4 \pi^2} \frac{\log (m_W^2/m_{Z'}^2)}{1-m_{Z'}^2/m_W^2}\,.
\end{eqnarray}
Interestingly, the sign of $\Delta$ from the ${\rm U(1)}_{L_\mu-L_\tau}$ interaction is consistent with that required by the difference between the experimental value and the SM prediction, Eq.~(\ref{Difference_tau}).
In Fig.~\ref{Fig-result}, the upper region of an orange curve is excluded by the $\tau \to \mu \nu_\tau \overline{\nu}_\mu$ decay limit at the $2 \sigma$ level.

\subsection{Neutrino trident production}

\begin{figure}
\begin{center}
\includegraphics[width=0.41\linewidth]{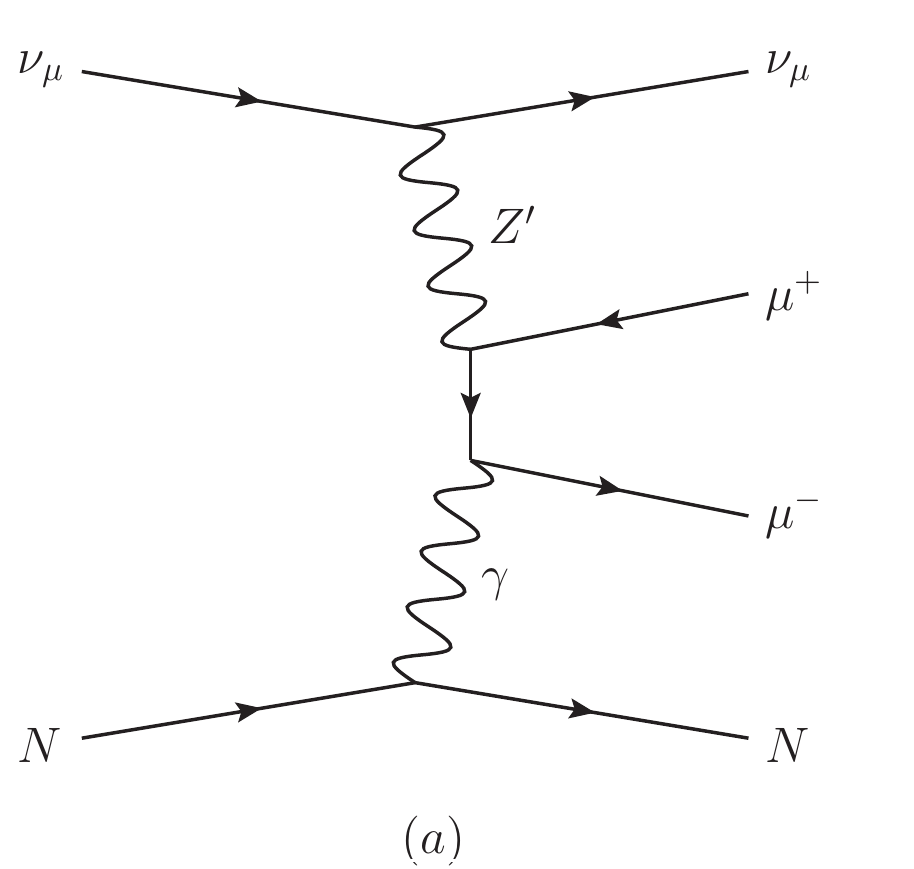}
\hspace*{0.5cm}
\includegraphics[width=0.42\linewidth]{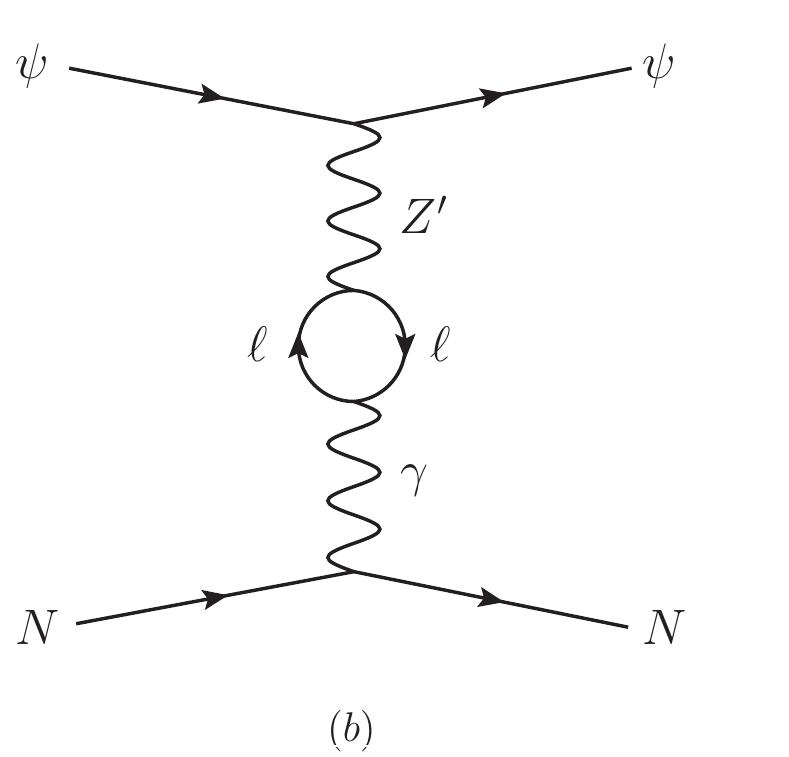}
\end{center}
\vspace*{-0.7cm}
\caption{(a) The leading order diagram with a $Z'$ exchange contributing to neutrino trident production. (b) Diagram of dominant direct detection process.
}
\label{Fig-NuDD}
\end{figure}

Neutrino trident production, $\nu_\mu N \to \nu_\mu N \mu^+ \mu^-$, has been observed by several neutrino beam experiments such as CHARM-II~\cite{Geiregat:1990gz} and CCFR~\cite{Mishra:1991bv}:
\begin{eqnarray}\label{TridentExp}
\frac{\sigma_{\rm CHARM-II}}{\sigma_{\rm SM}} &=& 1.58 \pm 0.57\,,\\
\frac{\sigma_{\rm CCFR}}{\sigma_{\rm SM}} &=& 0.82 \pm 0.28\,.
\end{eqnarray}
The measured cross sections are consistent with the SM prediction so that  stringently constrain our model.
In the SM, the neutrino trident production is induced by a $\mu^+\mu^-$ pair production from the scattering of a muon-neutrino in the Coulomb field of a target nucleus \cite{Altmannshofer:2014cfa, Altmannshofer:2014pba}.
In our model, the leading order correction is coming from the contribution of $Z'$ boson shown in Fig.~\ref{Fig-NuDD}(a) that interferes with the SM contribution from similar diagrams with a $W/Z$ boson exchange instead of the $Z'$.
In our analysis, we use the exclusion limit ($95\%$ C.L.) obtained from the CCFR data in Ref.~\cite{Altmannshofer:2014pba} which is shown as a light cyan-shaded region with the cyan dot-dashed curves in Fig.~\ref{Fig-result}.

%
%

\subsection{Dark matter direct detection}

DM direct detection experiments search for the recoil energy of nucleus by DM scattering off nucleus.
In this model, DM does not directly couple to quarks at tree-level.
However, one-loop suppressed scattering processes such as the one shown in Fig.~\ref{Fig-NuDD}(b) can still provide a sizable DM-nucleus scattering cross section in spite of the loop suppression factor.
The one-loop suppressed DM-nucleus scattering cross section is given by \cite{Bell:2014tta}
\begin{eqnarray}\label{DD}
\sigma_{\psi N} = \frac{\mu^2_N}{9\pi} \left[ \left( \frac{\alpha_{\rm EM} Z}{\pi \Lambda^2} \right) \log\left( \frac{m^2_\tau}{m^2_\mu} \right ) \right ]^2\,,
\end{eqnarray}
where $\Lambda = m_{Z'}/(g' \sqrt{Q'_\psi})$ is the effective cut-off scale, $\mu_N = m_N m_\psi/(m_N+ m_\psi)$ is the DM-nucleus reduced mass, and $Z$ is the atomic number, i.e. the EM charge of the target nucleus.
Note that Eq.~(\ref{DD}) originally has a log dependence on the renormalization scale due to the fermion loop.
However, such log dependences from $\mu$- and $\tau$-loops cancel each other out thanks to the relative sign difference between $\mu$- and $\tau$-loop induced diagrams.
In order to directly compare the DM-nucleus cross section with experimental bounds, we convert Eq.~(\ref{DD}) into the DM-nucleon cross section using the following relation:
\begin{eqnarray}
\sigma_{\psi n} = \frac{1}{A^2} \frac{\mu^2_n}{\mu^2_N}\sigma_{\psi N}\,,
\end{eqnarray}
where $A$ is the atomic mass number of the target nucleus and $\mu_n$ is the DM-nucleon reduced mass.
For $m_\psi \simeq 10~{\rm GeV}$, the most stringent direct detection bound is currently provided by the LUX experiment~\cite{Akerib:2013tjd}.
The LUX limit is shown as a purple dashed line in Fig.~\ref{Fig-result}.

\subsection{Searches for $Z \to 4\ell$ at the LHC and LEP}

\begin{figure}[h]
	\begin{center}
		\includegraphics[width=0.60\linewidth]{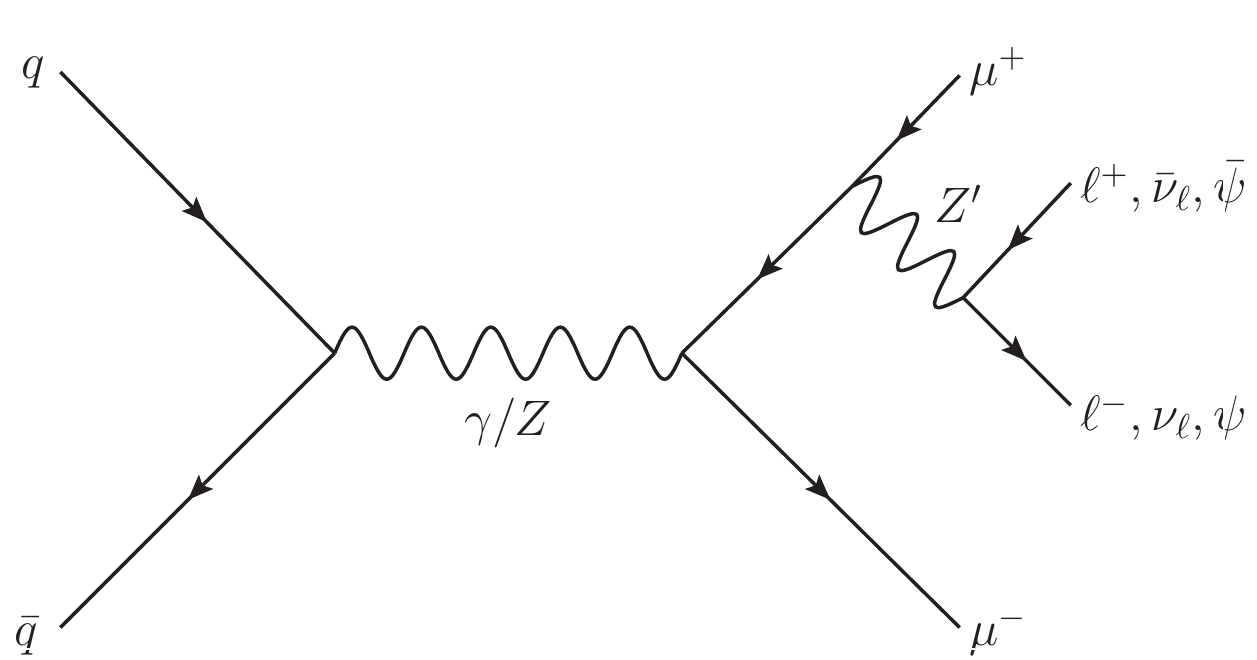}
	\end{center}
	\vspace*{-0.7cm}
	\caption{Feynman diagram for a $Z'$ boson production process at a hadron collider.
The $Z'$ boson is radiated from a lepton, and then decays into a pair of leptons or DM's.
}
\label{Fig-Production}
\end{figure}

The LHC results also provide constraints on the gauged lepton number interactions through the lepton productions.
A single $Z'$ production is allowed at tree-level at hadron colliders such as the LHC in $pp \to \ell^+\ell^- Z'$ where the $Z'$ boson is radiated from a lepton in the Drell-Yan process as shown in Fig.~\ref{Fig-Production} even though $Z'$ interaction is lepton-specific.
The produced $Z'$ boson subsequently decays either to a pair of charged-leptons, neutrinos or DM particles:
\begin{eqnarray}
Z' \to \mu^+ \mu^-, ~\tau^+ \tau^-, ~\nu_\ell\overline{\nu}_\ell, ~\psi \overline{\psi}\,,
\end{eqnarray}
if kinematically allowed.
These processes can be probed by detecting either one charged-lepton pair plus missing $E_T$ events or two charged-lepton pairs, i.e. $4\ell$, at the LHC.
In this work, we focus on the $4\ell$ signals due to its clean and distinctive signature.
If $m_{Z'} \gg (m_\tau, ~m_\psi)$, the branching ratios of the $Z'$ become
\begin{eqnarray}
{\rm Br}(Z' \to \ell\overline{\ell}) = {\rm Br}(Z' \to \nu_\ell\overline{\nu}_\ell)= {\rm Br}(Z' \to \psi \overline{\psi})/Q'^2_{\psi}\,.
\end{eqnarray}

%
%

\begin{figure}[h]
	\begin{center}
		\includegraphics[width=0.70\linewidth]{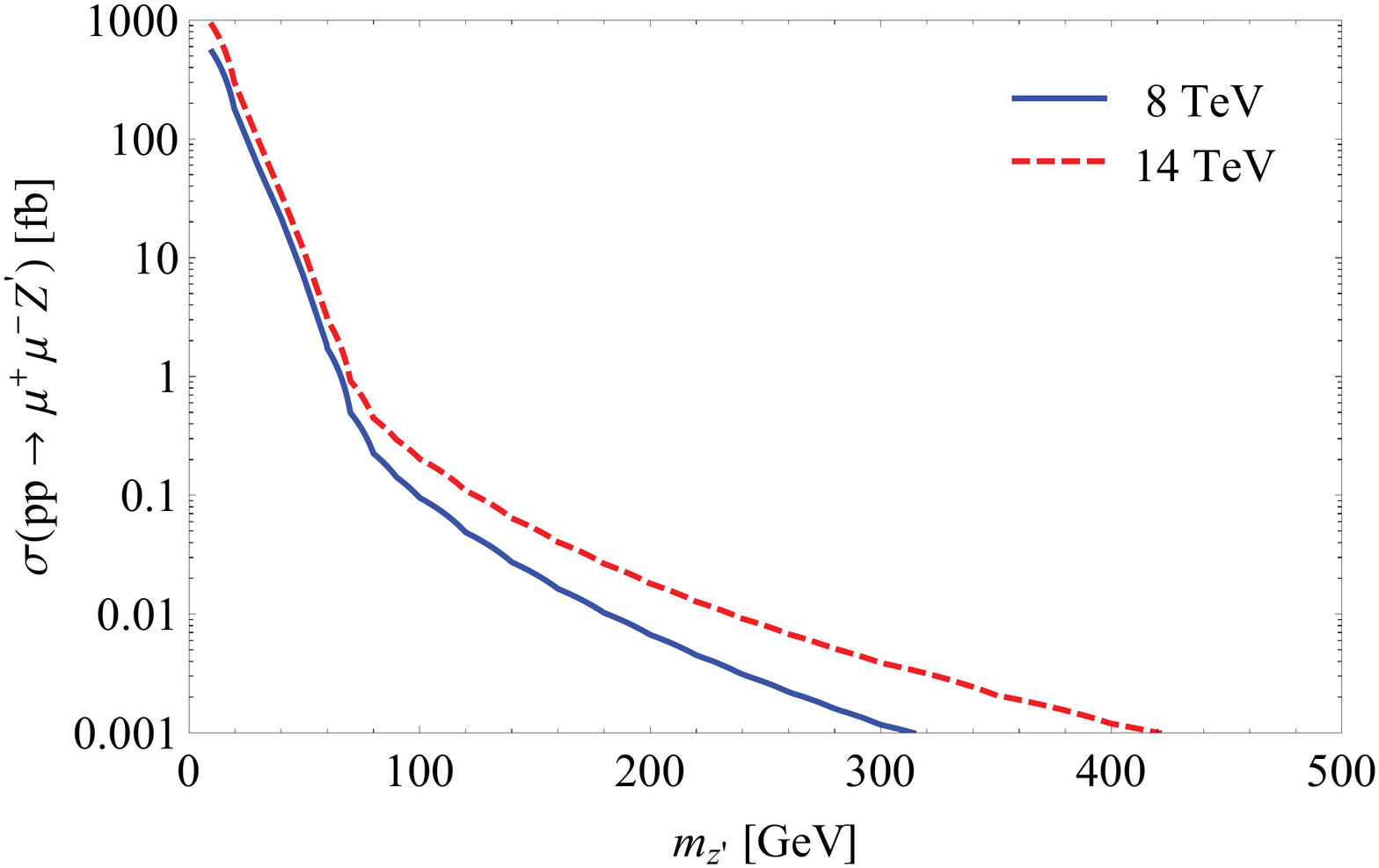}
	\end{center}
	\vspace*{-0.7cm}
	\caption{$Z'$ production cross section at the 8 TeV and 14 TeV LHC, through $pp\to\mu^-\mu^+ Z'$. We have set $g'=0.1$.
	}
\label{Fig-LHC}
\end{figure}

The leptophilic $Z'$ can be detected at the LHC in four charged-lepton final states.
The dominant SM backgrounds for this process are
\begin{eqnarray}
pp &\to& \ell^+\ell^- Z \to \ell^+\ell^-\ell^+\ell^-\,,\\
pp &\to& Z Z \to \ell^+\ell^-\ell^+\ell^-\,.
\end{eqnarray}
Four charged-lepton ($4\ell$) production at the $Z$ resonance has been already measured by ATLAS~\cite{Aad:2014wra} and CMS~\cite{CMS:2012bw} collaborations at the LHC.
Three final states have been well observed: $pp \to 4e,~2e2\mu,~4\mu$.
We consider only the four muon final state since in our scenario the $Z'$ does not couple to electrons.
In this analysis, we use the following selection cuts which is used in the ATLAS analysis~\cite{Aad:2014wra}:
\begin{itemize}
	\item[$\bullet $] $P_{T, \mu} > 4$ GeV and $\vert\eta\vert < 2.7$ for individual muons,
	\item[$\bullet $] Separation of muons: $\Delta R_{\mu\mu} > 0.1$,
	\item[$\bullet $] Invariant mass of a muon pair: $M_{\mu^+ \mu^-} > 5$ GeV,
	\item[$\bullet $] Invariant mass of four muons: 80 GeV $< m_{4\mu} <$ 100 GeV.
\end{itemize}
In Fig.~\ref{Fig-LHC}, we present the $Z'$ production cross sections through the $pp\to\mu^-\mu^+ Z'$ process for $g'=0.1$ at the 8 and 14 TeV LHC which is obtained using MadGraph~\cite{madgraph}.

%
%

%
\begin{figure}
\begin{center}
\includegraphics[width=0.49\linewidth]{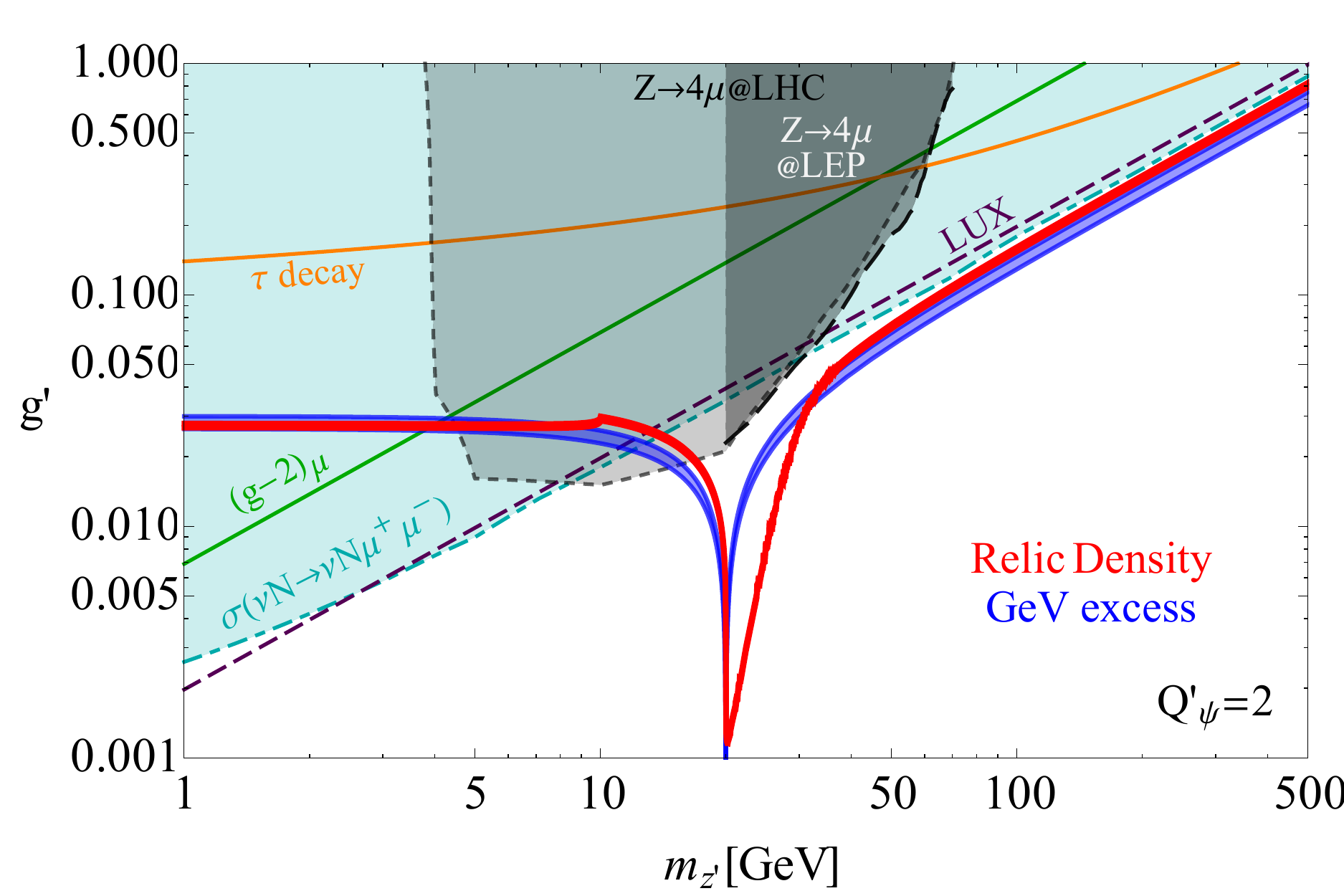}
\hspace*{0.1cm}
\includegraphics[width=0.49\linewidth]{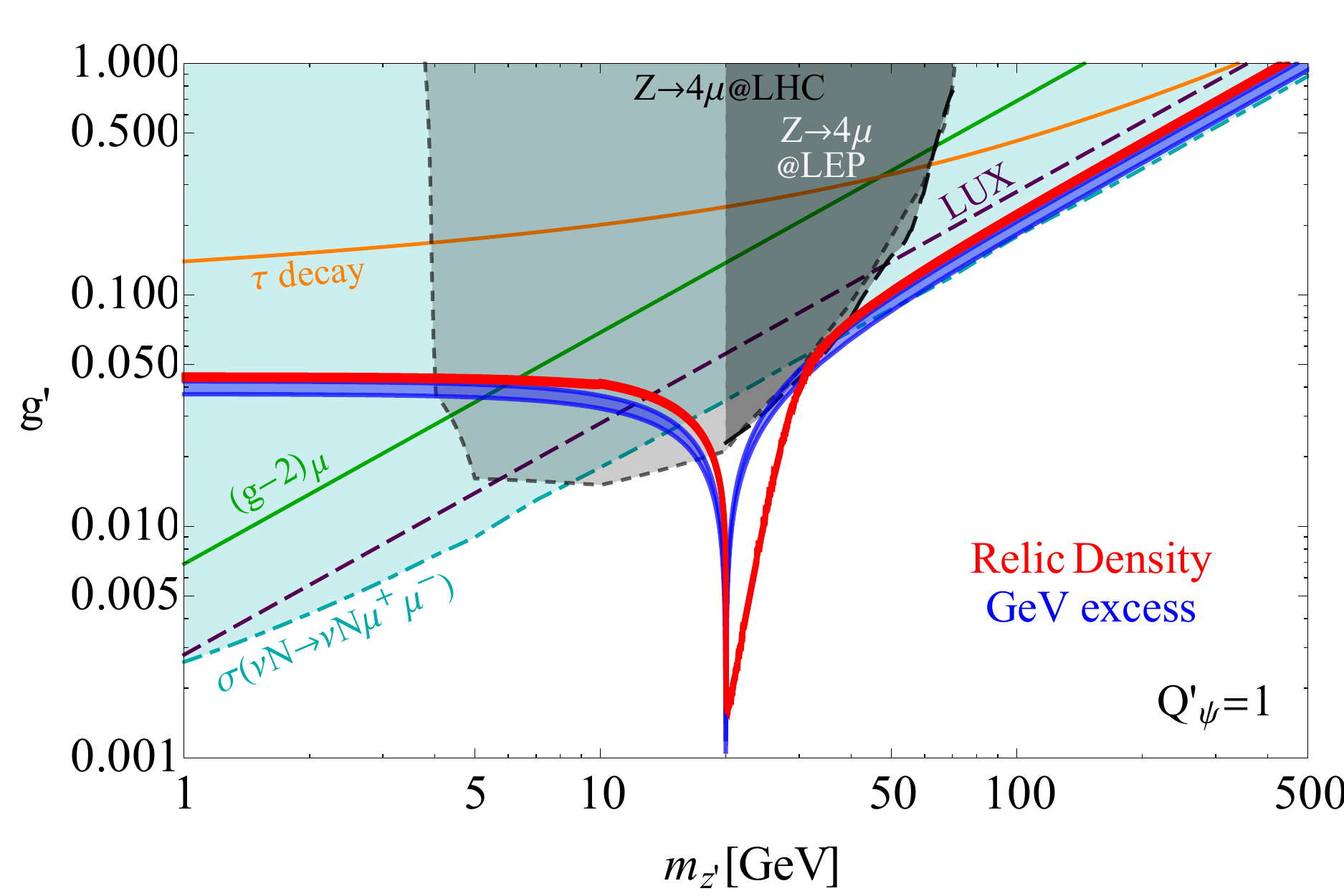}\\
\vspace*{0.3cm}
\includegraphics[width=0.49\linewidth]{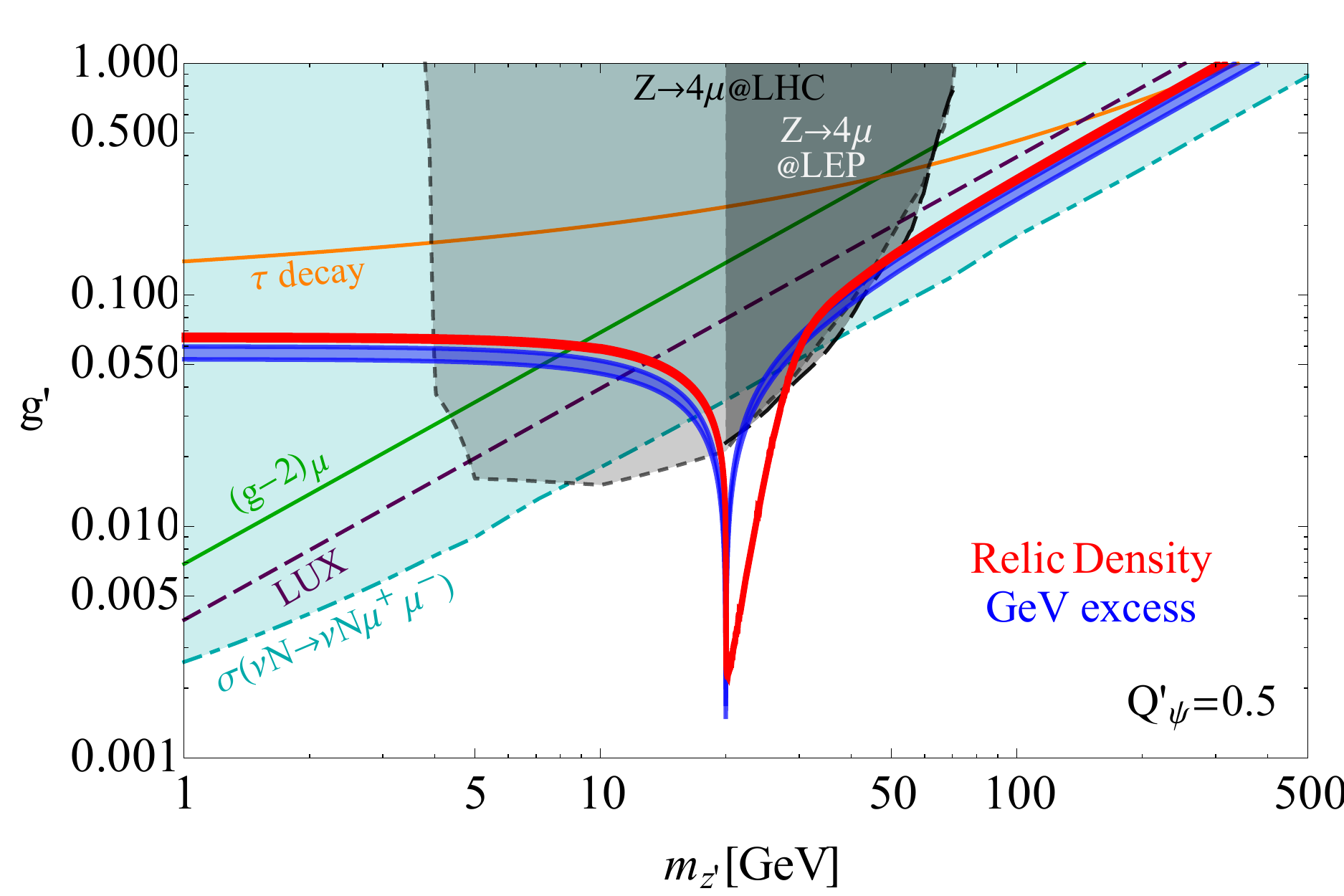}
\hspace*{0.1cm}
\includegraphics[width=0.49\linewidth]{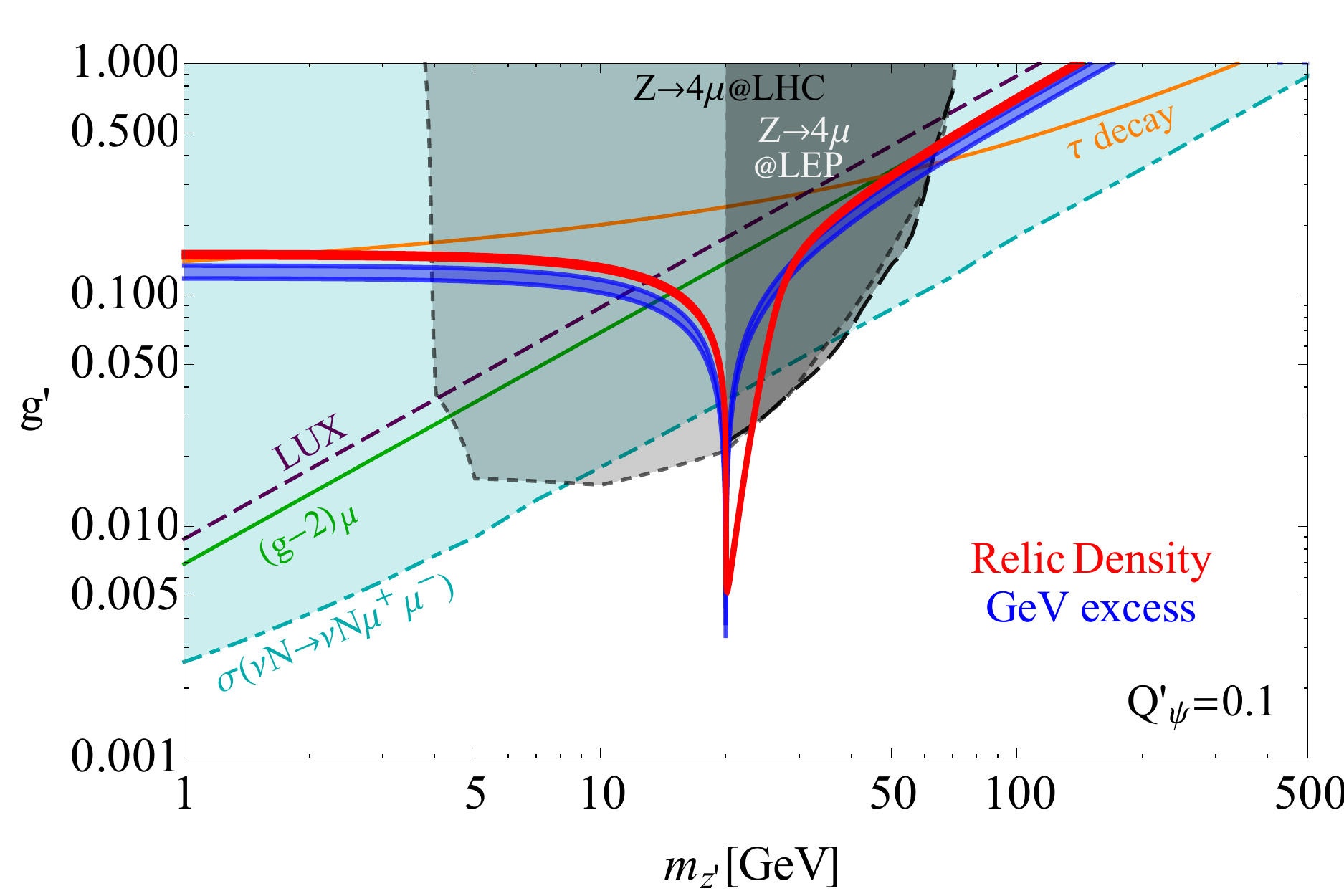}
\end{center}
\vspace*{-0.7cm}
\caption{Allowed parameter space of the U(1)$_{L_\mu-L_\tau}$ charged dark matter model in the $m_{Z'}-g'$ plane for four representative DM charges $Q'_\psi=$ 2, 1, 0.5, and 0.1 (from top-left to bottom-right), respectively.
We present the regions satisfying the DM relic abundance $0.11 < \Omega_{\rm DM}h^2 < 0.13$ and the annihilation cross section $\langle\sigma v \rangle_{\psi\overline{\psi}\to \mu^+\mu^-, \tau^+\tau^-} \approx (0.95-1.49) \times 10^{-26} {\rm cm}^3/{\rm s}$ required to fit the GC GeV excess as red and blue bands.
The upper regions of green, orange, cyan dot-dashed, purple dashed, gray dotted, and black long-dashed curves are constrained by $(g-2)_\mu$, $\tau$ decay, neutrino trident production, LUX, LHC, and LEP, respectively.
In this analysis, DM mass is fixed as $m_\psi=10$ GeV.
}
\label{Fig-result}
\end{figure}
%

This $Z'$ can be produced at tree-level at lepton colliders such as LEP through the similar process as shown in Fig.~\ref{Fig-Production} just by replacing $q\bar{q}$ with $e^+e^-$ since the gauged U(1)$_{L_\mu - L_\tau}$ boson also has no direct coupling to $e^+$ and $e^-$.
The potential constraint from LEP for $m_{Z'} < m_Z$ has been well studied in Ref.~\cite{Ma:2001md} through the process, $e^+e^- \to \mu^+\mu^- Z'$.
Despite much smaller total integrated luminosity, the limit from LEP is comparable to that from the 8 TeV LHC due to much cleaner signals.
In Fig.~\ref{Fig-result}, we present the LEP limit on $Z'$ from Ref.~\cite{Ma:2001md} as a dark-gray shaded region with the black long-dashed curves.

\subsection{Summary of experimental constraints on the ${\rm U}(1)_{L_\mu-L_\tau}$ model}

In Fig.~\ref{Fig-result}, we collectively depict all the relevant constraints to the gauged lepton number interaction in $m_{Z'} - g'$ plane.
\begin{itemize}
\item{\bf The limit from $(g-2)_\mu$}: We plot the $2\sigma$ limit from $(g-2)_\mu$ as a green solid line in the $m_{Z'}-g'$ plane for representative choices $Q'_\psi = 2$, 1, 0.5, and 0.1. The upper region of the green line is constrained by the current measurements of $(g-2)_\mu$.
\item{\bf The limit from $\tau \to \mu \nu_\tau \overline{\nu}_\mu$ decay}: The upper region of the orange curve is excluded by the $\tau \to \mu \nu_\tau \overline{\nu}_\mu$ decay limit at the $2\sigma$ level.
\item{\bf The limit from Neutrino trident production}: The $95\%$ C.L. exclusion limit
is shown as a light cyan-shaded region with a cyan dot-dashed curve.
\item{\bf The limit from dark matter direct detection}: The LUX limit is plotted as purple dashed lines for four representative values $Q'_\psi=2$, 1, 0.5, and 0.1.
\item{\bf LHC $Z\to 4\ell$ limit}: The light gray-shaded region with the gray dotted curve is excluded by measurements of the $Z \to 4\mu$ at the LHC~\cite{Altmannshofer:2014pba, Bell:2014tta}. The $Z \to 4\mu$ searches at the LHC strongly constrain the parameter space of $m_{Z'} \approx 5 - 40~{\rm GeV}$ since the $4\ell$ production has been measured at the $Z$ resonance and the selection cuts of $P_{T, \mu} > 4$ GeV and $M_{\mu^+ \mu^-} > 5$ GeV are used.
\item{\bf LEP $Z\to 4\ell$ limit}: Dark grad-shaded region with the black long-dashed curve is excluded by measurements of the $Z \to 4\mu$ at LEP~\cite{Ma:2001md}.
\end{itemize}
For $Q'_\psi \lesssim 1$, considerable parameter space has already been ruled out by neutrino trident production and $Z \to 4\mu$ observations at the 8 TeV LHC and LEP, except the region around the resonance of $m_{Z'} \approx 2m_\psi$.
In near future, for larger $Q'_\psi \gtrsim 1$ most of preferred parameter space will be verified by DM direct detection experiments such as XENON1T.
The region around the resonance will be complementarily proved by $Z \to 4\mu$ searches at the 14 TeV LHC.

\section{Conclusion}

In this work, we have explored a leptophilic DM model with the gauged U(1)$_{L_\mu-L_\tau}$ symmetry in the light of the Fermi-LAT GeV gamma-ray excess.
With this simple leptophilic DM model, we can simultaneously explain the observed DM relic abundance and the Fermi-LAT GeV excess.
Our leptophilic $Z'$ DM model additionally contributes to the muon $(g-2)$, tau decay process, and neutrino trident production.
In particular, neutrino trident production measurements provide the most stringent constraint to the DM model in most of the parameter space.
Despite the absence of direct couplings with quarks, this model can be strongly constrained by DM direct detection bounds through the loop-suppressed process.
For DM with a large charge under the U(1)$_{L_\mu-L_\tau}$, $Q'_\psi \gtrsim 2$, the current LUX direct search limit is comparable or stronger than the neutrino trident production limit.
The U(1)$_{L_\mu-L_\tau}$ gauge boson can be produced through the radiation process from Drell-Yan leptons, which has been constrained by $Z \to 4\mu$ searches at the LHC and LEP, especially for $m_{Z'} \approx 5 - 40~{\rm GeV}$.

\vspace{0.5 cm}
\begin{acknowledgements}
JCP and SCP are supported by the Basic Science Research Program through the National Research Foundation of Korea funded by the Ministry of Education, Science and Technology (based on contract numbers NRF-2013R1A1A2061561 and 2013R1A1A2064120, respectively).
\end{acknowledgements}


\begin{thebibliography}{99}

\bibitem{Jungman:1995df}
  G.~Jungman, M.~Kamionkowski and K.~Griest,
  Phys.\ Rept.\  {\bf 267}, 195 (1996)
  [hep-ph/9506380];
  G.~Bertone, D.~Hooper and J.~Silk,
  Phys.\ Rept.\  {\bf 405}, 279 (2005)
  [hep-ph/0404175].




\bibitem{Adriani:2008zr}
  O.~Adriani {\it et al.}  [PAMELA Collaboration],
  Nature {\bf 458}, 607 (2009)
  [arXiv:0810.4995 [astro-ph]];
  O.~Adriani {\it et al.}  [PAMELA Collaboration],
  Phys.\ Rev.\ Lett.\  {\bf 111}, 081102 (2013)
  [arXiv:1308.0133 [astro-ph.HE]].



\bibitem{Aguilar:2013qda}
  M.~Aguilar {\it et al.}  [AMS Collaboration],
  Phys.\ Rev.\ Lett.\  {\bf 110}, 141102 (2013);
  L.~Accardo {\it et al.}  [AMS Collaboration],
  Phys.\ Rev.\ Lett.\  {\bf 113}, 121101 (2014).


\bibitem{FermiLAT:2011ab}
  M.~Ackermann {\it et al.}  [Fermi-LAT Collaboration],
  Phys.\ Rev.\ Lett.\  {\bf 108}, 011103 (2012)
  [arXiv:1109.0521 [astro-ph.HE]].





\bibitem{Hooper1}
  L.~Goodenough and D.~Hooper,
  arXiv:0910.2998 [hep-ph].

\bibitem{Hooper2}
  D.~Hooper and L.~Goodenough,
  Phys.\ Lett.\ B {\bf 697}, 412 (2011)
  [arXiv:1010.2752 [hep-ph]].

\bibitem{Hooper3}
  D.~Hooper and T.~Linden,
  Phys.\ Rev.\ D {\bf 84}, 123005 (2011)
  [arXiv:1110.0006 [astro-ph.HE]].

\bibitem{Abazajian:2012pn}
  K.~N.~Abazajian and M.~Kaplinghat,
  Phys.\ Rev.\ D {\bf 86}, 083511 (2012)
  [arXiv:1207.6047 [astro-ph.HE]].

\bibitem{Hooper4}
  D.~Hooper and T.~R.~Slatyer,
  Phys.\ Dark Univ.\  {\bf 2}, 118 (2013)
  [arXiv:1302.6589 [astro-ph.HE]].

\bibitem{Gordon:2013vta}
  C.~Gordon and O.~Macias,
  Phys.\ Rev.\ D {\bf 88}, 083521 (2013)
  [arXiv:1306.5725 [astro-ph.HE]].

\bibitem{Huang:2013pda}
  W.~-C.~Huang, A.~Urbano and W.~Xue,
  arXiv:1307.6862 [hep-ph].

\bibitem{Abazajian:2014fta}
  K.~N.~Abazajian, N.~Canac, S.~Horiuchi and M.~Kaplinghat,
  arXiv:1402.4090 [astro-ph.HE].

\bibitem{HooperNew}
  T.~Daylan, D.~P.~Finkbeiner, D.~Hooper, T.~Linden, S.~K.~N.~Portillo, N.~L.~Rodd and T.~R.~Slatyer,
  arXiv:1402.6703 [astro-ph.HE].

\bibitem{Silk}
  T.~Lacroix, C.~Boehm and J.~Silk,
  arXiv:1403.1987 [astro-ph.HE].


\bibitem{Calore:2014xka}
  F.~Calore, I.~Cholis and C.~Weniger,
  arXiv:1409.0042 [astro-ph.CO].

\bibitem{Calore:2014nla}
  F.~Calore, I.~Cholis, C.~McCabe and C.~Weniger,
  arXiv:1411.4647 [hep-ph].

\bibitem{Kim:2015usa}
  D.~Kim and J.~C.~Park,
  arXiv:1507.07922 [hep-ph];
  D.~Kim and J.~C.~Park,
  arXiv:1508.06640 [hep-ph].


\bibitem{Belanger:2011ww}
  G.~Belanger and J.~C.~Park,
  JCAP {\bf 1203}, 038 (2012)
  [arXiv:1112.4491 [hep-ph]].


\bibitem{AMS2015}
  AMS-02 Collaboration,
  Talks at the `AMS Days at CERN', 15-17 April, 2015.

\bibitem{Giesen:2015ufa}
  G.~Giesen, M.~Boudaud, Y.~Genolini, V.~Poulin, M.~Cirelli, P.~Salati and P.~D.~Serpico,
  arXiv:1504.04276 [astro-ph.HE].



\bibitem{Okada:2013bna}
  N.~Okada and O.~Seto,
  Phys.\ Rev.\ D {\bf 89}, no. 4, 043525 (2014)
  [arXiv:1310.5991 [hep-ph]].

\bibitem{Alves:2014yha}
  A.~Alves, S.~Profumo, F.~S.~Queiroz and W.~Shepherd,
  Phys.\ Rev.\ D {\bf 90}, no. 11, 115003 (2014)
  [arXiv:1403.5027 [hep-ph]].

\bibitem{Ipek:2014gua}
  S.~Ipek, D.~McKeen and A.~E.~Nelson,
  Phys.\ Rev.\ D {\bf 90}, no. 5, 055021 (2014)
  [arXiv:1404.3716 [hep-ph]].

\bibitem{Basak:2014sza}
  T.~Mondal and T.~Basak,
  Phys.\ Lett.\ B {\bf 744}, 208 (2015)
  [arXiv:1405.4877 [hep-ph]].



\bibitem{Chun:2008by}
  E.~J.~Chun and J.~C.~Park,
  JCAP {\bf 0902}, 026 (2009)
  [arXiv:0812.0308 [hep-ph]];
  E.~J.~Chun, J.~C.~Park and S.~Scopel,
  JCAP {\bf 1002}, 015 (2010)
  [arXiv:0911.5273 [hep-ph]];
  S.~C.~Park and J.~Shu,
  Phys.\ Rev.\ D {\bf 79}, 091702 (2009)
  [arXiv:0901.0720 [hep-ph]];
  C.~R.~Chen, M.~M.~Nojiri, S.~C.~Park, J.~Shu and M.~Takeuchi,
  JHEP {\bf 0909}, 078 (2009)
  [arXiv:0903.1971 [hep-ph]].


\bibitem{Adriani:2010rc}
  O.~Adriani {\it et al.}  [PAMELA Collaboration],
  Phys.\ Rev.\ Lett.\  {\bf 105}, 121101 (2010)
  [arXiv:1007.0821 [astro-ph.HE]].




\bibitem{He:1990pn}
  X.~G.~He, G.~C.~Joshi, H.~Lew and R.~R.~Volkas,
  Phys.\ Rev.\ D {\bf 43}, 22 (1991).

\bibitem{He:1991qd}
  X.~G.~He, G.~C.~Joshi, H.~Lew and R.~R.~Volkas,
  Phys.\ Rev.\ D {\bf 44}, 2118 (1991).

\bibitem{Foot:1990mn}
  R.~Foot,
  Mod.\ Phys.\ Lett.\ A {\bf 6}, 527 (1991).


\bibitem{Baek:2008nz}
  S.~Baek and P.~Ko,
  JCAP {\bf 0910}, 011 (2009)
  [arXiv:0811.1646 [hep-ph]].

\bibitem{Bi:2009uj}
  X.~J.~Bi, X.~G.~He and Q.~Yuan,
  Phys.\ Lett.\ B {\bf 678}, 168 (2009)
  [arXiv:0903.0122 [hep-ph]].

\bibitem{Das:2013jca}
  M.~Das and S.~Mohanty,
  Phys.\ Rev.\ D {\bf 89}, no. 2, 025004 (2014)
  [arXiv:1306.4505 [hep-ph]].


\bibitem{Kong:2014haa}
  K.~Kong and J.~C.~Park,
  Nucl.\ Phys.\ B {\bf 888}, 154 (2014)
  [arXiv:1404.3741 [hep-ph]].


\bibitem{Abazajian:2010zy}
  K.~N.~Abazajian,
  JCAP {\bf 1103}, 010 (2011)
  [arXiv:1011.4275 [astro-ph.HE]].

\bibitem{Hooper:2013nhl}
  D.~Hooper, I.~Cholis, T.~Linden, J.~Siegal-Gaskins and T.~Slatyer,
  Phys.\ Rev.\ D {\bf 88}, 083009 (2013)
  [arXiv:1305.0830 [astro-ph.HE]].


\bibitem{Linden:2012iv}
  T.~Linden, E.~Lovegrove and S.~Profumo,
  Astrophys.\ J.\  {\bf 753}, 41 (2012)
  [arXiv:1203.3539 [astro-ph.HE]].

\bibitem{Macias:2013vya}
  O.~Macias and C.~Gordon,
  Phys.\ Rev.\ D {\bf 89}, 063515 (2014)
  [arXiv:1312.6671 [astro-ph.HE]].





\bibitem{Kyae:2013qna}
  B.~Kyae and J.~C.~Park,
  Phys.\ Lett.\ B {\bf 732}, 373 (2014)
  [arXiv:1310.2284 [hep-ph]].


\bibitem{Planck2015}
  P.~A.~R.~Ade {\it et al.}  [Planck Collaboration],
  [arXiv:1502.01589 [astro-ph.CO]].

\bibitem{Griest:1990kh}
  K.~Griest and D.~Seckel,
  Phys.\ Rev.\ D {\bf 43}, 3191 (1991).



\bibitem{Baek:2001kca}
  S.~Baek, N.~G.~Deshpande, X.~G.~He and P.~Ko,
  Phys.\ Rev.\ D {\bf 64}, 055006 (2001)
  [hep-ph/0104141].



\bibitem{Ma:2001md}
  E.~Ma, D.~P.~Roy and S.~Roy,
  Phys.\ Lett.\ B {\bf 525}, 101 (2002)
  [hep-ph/0110146].

\bibitem{Harigaya:2013twa}
  K.~Harigaya, T.~Igari, M.~M.~Nojiri, M.~Takeuchi and K.~Tobe,
  JHEP {\bf 1403}, 105 (2014)
  [arXiv:1311.0870 [hep-ph]].

\bibitem{Agashe:2014kda}
  K.~A.~Olive {\it et al.}  [Particle Data Group Collaboration],
  Chin.\ Phys.\ C {\bf 38}, 090001 (2014).


\bibitem{Pich:2013lsa}
  A.~Pich,
  Prog.\ Part.\ Nucl.\ Phys.\  {\bf 75}, 41 (2014)
  [arXiv:1310.7922 [hep-ph]].

\bibitem{Altmannshofer:2014cfa}
  W.~Altmannshofer, S.~Gori, M.~Pospelov and I.~Yavin,
  Phys.\ Rev.\ D {\bf 89}, no. 9, 095033 (2014)
  [arXiv:1403.1269 [hep-ph]].


\bibitem{Geiregat:1990gz}
  D.~Geiregat {\it et al.}  [CHARM-II Collaboration],
  Phys.\ Lett.\ B {\bf 245}, 271 (1990).

\bibitem{Mishra:1991bv}
  S.~R.~Mishra {\it et al.}  [CCFR Collaboration],
  Phys.\ Rev.\ Lett.\  {\bf 66}, 3117 (1991).

\bibitem{Altmannshofer:2014pba}
  W.~Altmannshofer, S.~Gori, M.~Pospelov and I.~Yavin,
  Phys.\ Rev.\ Lett.\  {\bf 113}, 091801 (2014)
  [arXiv:1406.2332 [hep-ph]].


\bibitem{Bell:2014tta}
  N.~F.~Bell, Y.~Cai, R.~K.~Leane and A.~D.~Medina,
  Phys.\ Rev.\ D {\bf 90}, no. 3, 035027 (2014)
  [arXiv:1407.3001 [hep-ph]].

\bibitem{Akerib:2013tjd}
  D.~S.~Akerib {\it et al.}  [LUX Collaboration],
  Phys.\ Rev.\ Lett.\  {\bf 112}, 091303 (2014)
  [arXiv:1310.8214 [astro-ph.CO]].

\bibitem{Aad:2014wra}
  G.~Aad {\it et al.}  [ATLAS Collaboration],
  Phys.\ Rev.\ Lett.\  {\bf 112}, no. 23, 231806 (2014)
  [arXiv:1403.5657 [hep-ex]].

\bibitem{CMS:2012bw}
  S.~Chatrchyan {\it et al.}  [CMS Collaboration],
  JHEP {\bf 1212}, 034 (2012)
  [arXiv:1210.3844 [hep-ex]].



\bibitem{madgraph}
  J.~Alwall, M.~Herquet, F.~Maltoni, O.~Mattelaer, T.~Stelzer,
  JHEP {\bf 1106}, 128 (2011)
  [arXiv:1106.0522[hep-ph]]











\end{thebibliography}
\end{document}